\documentclass[letterpaper, 10 pt, conference]{ieeeconf}
\IEEEoverridecommandlockouts 
\usepackage{cite}
\usepackage{amsmath,amssymb,amsfonts}

\usepackage{graphicx}
\usepackage{float}
\usepackage{textcomp}
\usepackage{graphicx}
\usepackage{subfigure}
\usepackage{caption}
\usepackage{subcaption}
\usepackage{lipsum}
\usepackage{multicol}
\usepackage{booktabs}
\usepackage[colorlinks=true,linkcolor=blue,citecolor=blue,urlcolor=blue]{hyperref}

\newtheorem{theorem}{Theorem}
\newtheorem{lemma}{Lemma}

\newtheorem{remark}{Remark}
\newtheorem{assumption}{Assumption}
\usepackage{cuted}
\usepackage[english]{babel}
\usepackage{xcolor}
    
    \makeatletter
 \let\old@ps@headings\ps@headings
 \let\old@ps@IEEEtitlepagestyle\ps@IEEEtitlepagestyle
 \def\confheader#1{%
 \def\ps@headings{%
 \old@ps@headings%
 \def\@oddhead{\strut\hfill#1\hfill\strut}%
 \def\@evenhead{\strut\hfill#1\hfill\strut}%
 }%
 \def\ps@IEEEtitlepagestyle{%
 \old@ps@IEEEtitlepagestyle%
 \def\@oddhead{\strut\hfill#1\hfill\strut}%
 \def\@evenhead{\strut\hfill#1\hfill\strut}%
 }%
 \ps@headings%
 }
 \makeatother

 \usepackage[pscoord]{eso-pic}

  \makeatletter
\newcommand{\linebreakand}{%
  \end{@IEEEauthorhalign}
  \hfill\mbox{}\par
  \mbox{}\hfill\begin{@IEEEauthorhalign}
}
\makeatother  

  \newtheorem{corollary}{Corollary}
\begin{document}

\title{Model Reference Adaptive Control with Time-Varying State and Input Constraints
}

\author{Poulomee~Ghosh and Shubhendu~Bhasin
\thanks{Poulomee~Ghosh and Shubhendu~Bhasin are with Department of Electrical Engineering, Indian Institute of Technology Delhi, New Delhi, India. 
        {\tt\small (Email: Poulomee.Ghosh@ee.iitd.ac.in, sbhasin@ee.iitd.ac.in)}}}



\maketitle
\begin{abstract}
This paper presents a model reference adaptive control (MRAC) framework for uncertain linear time-invariant (LTI) systems subject to user-defined, time-varying state and input constraints. The proposed design seamlessly integrates a time-varying barrier Lyapunov function (TVBLF) to enforce state constraints with a time-varying saturation function to handle input limits. These time-varying constraints can be designed as performance functions to shape transient and steady-state behaviors for both state and input. A key contribution is the derivation of a verifiable, offline feasibility condition to check the existence of a valid control policy for a given set of constraints. To the best of our knowledge, this is the first adaptive control methodology to simultaneously handle both time-varying state and input constraints without resorting to online optimization. Simulation results validate the efficacy of the proposed constrained MRAC scheme.

\end{abstract}


\section{Introduction}
\label{sec:intro}
 Model reference adaptive control (MRAC) has long been recognized for its ability to handle uncertainties and ensure reference model tracking \cite{mrac3, mrac2, classMRAC}. However, classical MRAC frameworks, while guaranteeing asymptotic tracking and boundedness of closed-loop signals, do not inherently respect pre-specified constraints on system states and inputs. This gap is significant, as real-world safety-critical applications- from autonomous vehicles to aerospace systems, demand strict adherence to physical, operational, and safety limits. 

A prominent optimization-free approach for handling state constraints in adaptive control is through barrier Lyapunov functions (BLFs) \cite{BLF, BLF2, Lafflitto}. Unlike optimization-based methods like model predictive control (MPC) \cite{mpc11, mpc12} or control barrier function (CBF) \cite{cbf, taylor2020adaptive}, which often require solving complex optimization problems online, BLFs integrate constraints directly into the control design. The BLF grows to infinity as the constrained state approaches the boundary of the safe set, effectively creating a potential barrier that the controller prevents the system from crossing. While effective, most existing BLF-based adaptive control schemes are limited to constant, time-invariant constraint boundaries. This raises a critical question: \textit{Is it practical to impose uniform constraint bounds across transient and steady-state phases, considering the inherently larger tracking errors during transients?} Intuitively, a more flexible approach is desirable; one that permits larger transients initially and tightens the constraints as the system converges. This motivates the concept of time-varying BLFs (TVBLFs) \cite{tvblf1}, where the constraint boundary evolves dynamically over time. The key challenge, however, lies in defining how this evolving boundary should be shaped to reflect specific performance objectives. 
This brings attention to the concept of prescribed performance functions (PPFs), which are smooth user-defined curves that dictate how the tracking error should behave over time. By specifying an initial error bound, a final bound, and a convergence rate, the user can shape the error trajectory.
The idea of time-varying performance bounds is central to prescribed performance control (PPC) \cite{ppc1, ppc2} and funnel control (FC) \cite{fc1, fc2}, which shape the tracking error trajectory within user-defined, shrinking envelopes. However, these methods face challenges when hard input constraints are present. To maintain feasibility under actuator saturation, the PPC typically modifies the reference trajectory online \cite{ppc3}, while the FC relaxes the performance funnel \cite{funnelsaturation1, fc3}, both compromising the original control objective. 
In our earlier work, we proposed a state and input constrained MRAC \cite{ghosh2022state}, ensuring that the trajectory tracking error remains within fixed user-defined bounds. Although this ensured constraint satisfaction, the use of fixed constraints lacked the flexibility of dynamic performance adaptation and could become overly conservative for systems with varying operational limits. 

\subsection{Contributions}
This paper addresses the critical, yet unsolved, challenge of designing adaptive controllers that can handle simultaneously occurring time-varying state and input constraints. This problem is highly relevant in systems where actuator capabilities may degrade over time (e.g., due to battery depletion, thermal derating, etc.) or where it is desirable to impose a time-varying input budget that aligns with the convergence of the system state to mitigate overshoot and actuator fatigue.
The contributions of this work are threefold:
\begin{enumerate}
    \item The proposed MRAC design integrates a TVBLF for time-varying state constraints and a saturated controller for time-varying input constraints within a unified, optimization-free architecture.    
    \item We derive a sufficient condition that can be verified offline to check the existence of a feasible control policy for any given pair of time-varying state and input constraints, thus avoiding the computational burden of online feasibility monitoring. 
    \item We show that by designing the time-varying bounds as PPFs, our approach handles dynamically shrinking error and input envelopes, while preserving the original reference trajectory and user-defined constraint boundaries during saturation.
\end{enumerate}

\subsection{Notations}
Throughout this paper $\mathbb{R}$ denotes the set of real numbers, $\mathbb{R}^{p \times q}$ denotes set of $p\times q$ real matrices, the identity matrix in $\mathbb{R}^{p \times p}$ is denoted by $I_{p}$ and $\|\cdot\|$ represents the Euclidean vector norm and corresponding induced matrix norm. The minimum and maximum eigen values of a matrix $M$ is denoted by $\lambda_{min}\{M\}$ and $\lambda_{max}\{M\}$, respectively. $\mathcal{N}(\mu,\Sigma)$ denotes a multivariate Gaussian distribution with mean vector $\mu$ and covariance matrix $\Sigma$, $\mathbf{1}\{\cdot\}$ is the indicator function that equals $1$ if the condition inside holds and $0$ otherwise and $\sup_{\underline{t}\le t \le \overline{t}}(\cdot)$ represents the least upper bound of $(\cdot)$ for $t\in[\underline{t},\overline{t}]$.

\section{Problem Formulation and Preliminaries}
\label{sec:LFSR}
\subsection{Problem Statement}

Consider a linear time-invariant system
\begin{align}
    \dot{x}=Ax+Bu
    \label{plant}
\end{align}
where $x(t)\in \mathbb{R}^n$ denotes the system state, $u(t) \in \mathbb{R}^m$ denotes the control input, $A \in \mathbb{R}^{n \times n}$ is the unknown system matrix, and  $B \in \mathbb{R}^{n \times m}$ is the input matrix, assumed to be full rank and known. The pair $(A,B)$ is assumed to be stabilizable. While the analysis can be extended to cases with an unknown input matrix $B$ \cite{nussbaum}, this scenario is not considered here to maintain a clear focus on the primary contribution: the handling of time-varying constraints. 
The system is subject to the following time-varying state and input constraints:
\begin{align}
    &\|x(t)\|< \phi_x(t) \quad \forall t\ge 0 
    \label{statecon}\\
    &\|u(t)\|\leq\phi_u(t) \quad \forall t\ge 0\label{incon}
\end{align}
where $\phi_x(t), \;\phi_u(t)\colon [0,\infty)\to\mathbb R^{+}$ are continuously differentiable, positive user-defined functions. 
A stable reference model is considered as
\begin{align}
    \dot{x}_r=A_rx_r+B_rr
    \label{ref}
\end{align}
where $x_r(t)\in \mathbb{R}^n$ is the reference model state and $r(t) \in \mathbb{R}^{m}$ is a  piecewise continuous reference input bounded such that $\sup_{t\ge0}\|r(t)\|<\bar{r}$, where $\bar{r}>0$ is a known constant. $A_r \in \mathbb{R}^{n \times n}$, $B_r \in \mathbb{R}^{n \times m}$ are known. It is assumed that $A_r$ is Hurwitz i.e. for every $Q=Q^{\top}>0$, there exists $P=P^{\top}>0$ such that
\begin{align}
   A_r^{\top}P+PA_r+Q=0 
   \label{lyapeq}
\end{align} 
The control objective is to design a feasible input $u(t)$, if it exists, such that $x(t)$ tracks $x_r(t)$ while simultaneously satisfying the pre-defined time-varying state and input constraints, as stated in (\ref{statecon}) and (\ref{incon}), respectively. 
The dynamics of the trajectory tracking error is defined as
\begin{align}
    e=x-x_r
    \label{eeq}
\end{align}
\begin{assumption}
\label{ref_assumption}
The reference model states are bounded such that $\|x_r(t)\|\leq{\mathcal{X}}_r(t)<\phi_x(t)$, where $\mathcal{X}_r(t):\mathbb{R}_{+}\rightarrow \mathbb{R}_{+}$ is a continously differentiable known function.
\end{assumption}
Using Assumption \ref{ref_assumption}, the constraint on the plant state can be transformed into the constraint on the trajectory tracking error, i.e. $\|e(t)\|<\phi_e(t) \implies \|x(t)\|<\phi_x(t)$ $\forall t \geq 0$, where $\phi_e(t)=\phi_x(t)-{\mathcal{X}}_r(t)$ is a known positive time-varying function.

\subsection{MRAC Design}

Consider the classical certainty equivalence adaptive control law 
\begin{align}
    u=\hat{K}_xx+{K}_rr
    \label{ueq}
\end{align}
where $K_x\in \mathbb{R}^{m\times n}$, $K_r\in\mathbb{R}^{m \times m}$ are controller parameters and $\hat{K}_x(t)\in \mathbb{R}^{m\times n}$ is the estimate $K_x$. 
\begin{assumption}
\label{matching_condition_assumption}
There exists controller parameters $K_x$ and $K_r$ such that the following matching conditions are satisfied.
\begin{subequations}
    \begin{equation}
        A+BK_x=A_r \label{mc11}
    \end{equation}
    \begin{equation}
         BK_r=B_r
    \label{mc12}
    \end{equation}
    \label{mc}
\end{subequations}
where the controller gains are assumed to be bounded, i.e., $\|K_x\|<\bar{K}_x$ and $\|K_r\|<\bar{K}_r$, with $\bar{K}_x,\bar{K}_r>0$ are known constants.
\end{assumption}
This is a structural assumption common in MRAC, implying that the reference model dynamics are achievable for the given plant. The condition $BK_r=B_r$ requires that the column space of $B_r$ be a subspace of the column space of $B$.
\begin{remark} 
Assuming a known bound on the ideal gain is standard in projection-based MRAC, as it ensures parameter evolution within a compact set. Although the exact $K_x$ is unknown, a conservative bound can often be estimated from physical knowledge of the system (e.g., limits on mass, inertia, or actuator capabilities), and this bound is then used to define the projection set.

\end{remark}

  Using \eqref{eeq}, (\ref{ueq}) and (\ref{mc}), the closed-loop error dynamics can be obtained as
\begin{align}
  \dot{e}=A_re+B\tilde{K}_xx+BK_rr
\end{align}
where $\tilde{K}_x(t)\triangleq \hat{K_x}(t)- K_x\in\mathbb{R}^{m \times n}$ denote the parameter estimation error. 

%
The classical adaptive update laws are given as \cite{classMRAC}
    \begin{equation}
        \dot{\hat{K}}_x=-{\Gamma_xB^{\top}Pex^{\top}} 
         \label{MRAC}
    \end{equation}
where, $\Gamma_x\in \mathbb{R}^{m \times m}$ is a  positive definite adaptation gain matrices. using Lyapunov analysis, we can prove that all the closed-loop signals remain bounded and the trajectory tracking error converges to zero asymptotically \cite{mrac2}, \cite{classMRAC}.\\

\section{Constrained MRAC Design}
\label{sec:methodology}
Our proposed methodology integrates two key components: a saturated control input to handle time-varying input constraints, and an adaptive law based on a TVBLF to enforce time-varying state constraints.
\subsection{Saturated Control Design}
\begin{figure}[H]
    \centering
\includegraphics[width=0.8\linewidth]{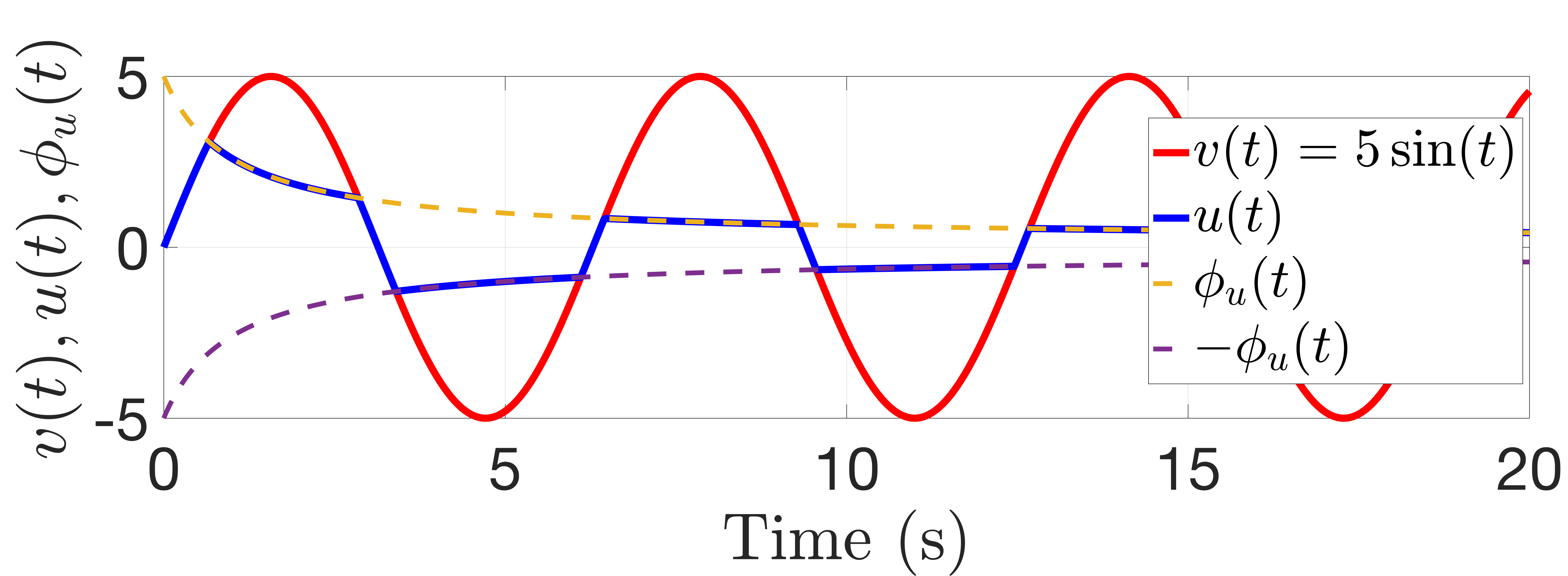}
    \caption{Saturated input using (\ref{sfc}), which satisfies (\ref{ppf}) with $\phi_u(t)=4.8\exp(-t)+0.2$.}
    \label{sat_u}
\end{figure}

To satisfy the time-varying input constraints, we define the following saturated input
\begin{align}
    u(t)
=\begin{cases}
v(t) & \|v(t)\|\le \phi_u(t) \\
\phi_u(t)\ \displaystyle\frac{v(t)}{\|v(t)\|} & \|v(t)\|>\phi_u(t)
\end{cases}
\label{sfc}
\end{align}
where $v(t)\in \mathbb{R}^m$ is an auxiliary control input, designed as
\begin{align}
&v=\hat{K}_x x+{K}_r r-\frac{\dot{\phi_e}}{\phi_e}B^{\dagger}e
\label{pc1}
\end{align}
 where the time derivative of $\phi_e(t)$ is denoted by $\dot{\phi}_e(t)=\frac{d}{dt}\phi_e(t) $ and $B^{\dagger}$ is the left inverse\footnote{The left inverse exists since $B$ is assumed to have full column rank.} of $B$, which satisfies \(B^{\dagger} = (B^{\top} B)^{-1} B^{\top}\). The saturated controller in (\ref{sfc}) ensures $\|u(t)\|<\phi_u(t)$ $\forall t \ge 0$, saturating any auxiliary input \(v(t)\) that would exceed the instantaneous actuator limit, as shown in Fig.~\ref{sat_u}.
Using (\ref{mc}) and (\ref{pc1})  the closed-loop error dynamics is obtained as
\begin{align}
\dot{e}=(A_r-\frac{\dot{\phi_e}}{\phi_e})e+B\tilde{K}_xx+B{K}_rr+B\Delta u
\end{align}
where `$\Delta u$' is the saturation deficiency defined as $\Delta u(t) := u(t) - v(t)$.\\
\subsection{TVBLF design}
To ensure the time-varying state constraint satisfaction, we design an adaptive controller employing a time-varying BLF, so that the evolution of the tracking error, and consequently the plant state is confined within a user-defined dynamic performance envelope.
\begin{figure}[H]
\centering
 \subfigure[]{\includegraphics[width=4.2cm]{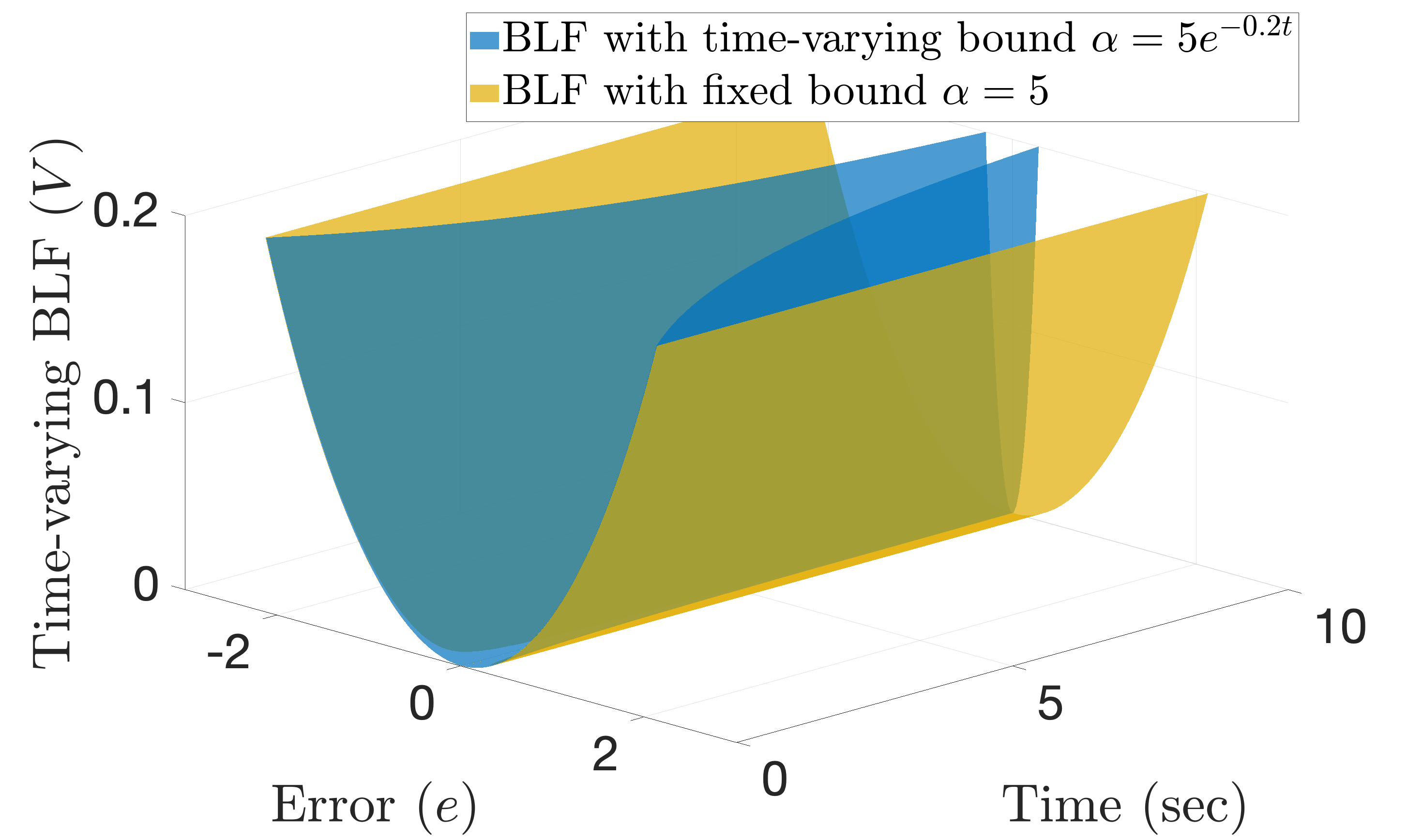}}
 \subfigure[]{\includegraphics[width=4.2cm]{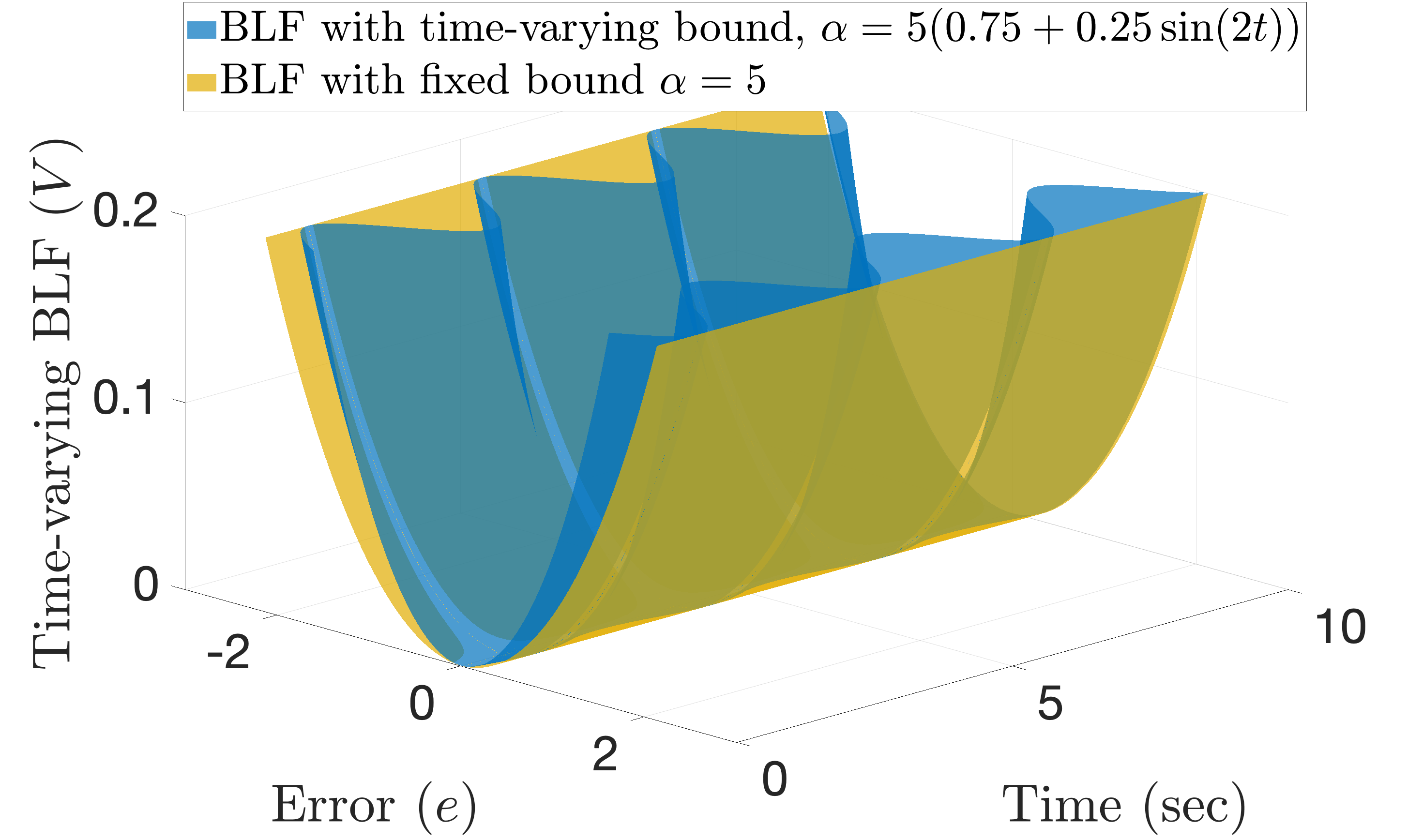}}
 \caption{BLF $\bigg(V=\log\frac{\alpha^2}{{\alpha}^2-e^2}\bigg)$ with fixed and time-varying bounds.}
 \end{figure}

To ensure the error remains within the time-varying bound $\|e(t)\|<\phi_e(t)$, we introduce the following Time-Varying Barrier Lyapunov Function (TVBLF):
\begin{equation}
    V_e(t) := \log \left( \frac{\phi^{'2}_e}{\phi^{'2}_e - e^{\top} P e} \right)
    \label{tvblf}
\end{equation}
defined on the set $\Omega^{'}_e(t)\triangleq\{e(t)\in\mathbb{R}^n: e(t)^{\top}Pe(t) < {\phi}_e^{'^2}(t)\}$,  where $\phi^{'}_e(t)=\phi_e(t)\sqrt{\lambda_{min}\{P\}}$.  The TVBLF in (\ref{tvblf}) imposes a dynamically evolving constraint on the error norm. Compared to BLF with fixed constraint, this provides additional flexibility to handle large initial transients without sacrificing steady-state accuracy.
%
\subsection{TVBLF Based Adaptive Update Law}
The adaptive update laws are defined as
\begin{equation}
\dot{\hat{K}}_x=\text{Proj}_{\Omega_1}\bigg(-\frac{\Gamma_xB^{\top}Pex^{\top}}{\phi_e^{'2}-e^{\top}Pe}\bigg)
    \label{updatelaws}
\end{equation}
 The projection operator \(\text{proj}_{*}(\cdot)\) \cite{lavretsky2012robust} ensures that the parameter update law \((\cdot)\) keeps the parameters bounded within a convex and compact region, denoted by `\(*\)', in the parameter space. In this context, the convex function associated with the projection operator is chosen as \(f(\hat{K}_x) = \|\hat{K}_x\|^2\) and the convex and compact region is defined by $\Omega_1=\{\hat{K}_x\in\mathbb{R}^{m\times n}|\|\hat{K}_x\|^2 \leq \bar{K}_x^2\}$, which align with Assumption \ref{matching_condition_assumption}. Furthermore, by invoking \cite[Lemma~6]{lavretsky2011projection}, it can be asserted that the projection operator introduces a negative semi-definite term in the Lyapunov derivative when the estimate is at the boundary of the convex and compact region, thereby preserving the stability results.
 \subsection{Feasibility and Stability Analysis}
 A key advantage of our approach is the ability to verify its feasibility before implementation. We establish a sufficient condition that checks the existence of a control policy that can satisfy both constraints simultaneously. 

\begin{theorem}
\label{maintheorem}
    Consider the LTI plant (\ref{plant}) and the reference model (\ref{ref}). Under Assumptions \ref{ref_assumption}-\ref{matching_condition_assumption}, the proposed controller (\ref{sfc}), (\ref{pc1}) with the adaptive update law (\ref{updatelaws}) ensures that  the state \eqref {statecon} and input \eqref{incon} constraints are satisfied for $t\ge 0$ and that all closed loop signals remain bounded provided the following feasibility condition C1 holds.\\
    \textbf{(C1):} The time-varying input constraint function satisfies the following inequality $\forall t \ge 0$:
        \begin{align}
{\phi}_u(t)>{\phi}_x(t) (\bar{K}_x-\eta)+ |\dot{\phi}_e(t)|\|B^{\dagger}\|+\eta \mathcal{X}_r(t) +\bar{K}_r \bar{r} \label{C11}
        \end{align}
        where $\eta=\frac{\lambda_{min}\{Q\}}{2\lambda_{max}\{P\}\|B\|}>0$ is a known constant.
\end{theorem}
\begin{proof}
Consider the candidate Lyapunov function as
    \begin{equation}
    V(e,\tilde{K}_x,t)
    = \frac{1}{2}V_e + \mathrm{tr} \left( \tilde{K}_x^{\top} \Gamma_x^{-1} \tilde{K}_x \right) 
    \label{lyap}
\end{equation}
where \( V_1 \) is the time-varying BLF defined in (\ref{tvblf}). Taking the time-derivative of $V$ along the system trajectory 
\begin{align}
\dot{V}=&\frac{\dot{e}^{\top}Pe+e^{\top}P\dot{e}}{2(\phi^{'2}_e-e^{\top}Pe)}+\frac{\dot{\phi^{'}_e}e^{\top}Pe}{\phi_e^{'}(\phi^{'2}_e-e^{\top}Pe)}+tr(\tilde{K}_x^{\top}\Gamma_x^{-1}\dot{\hat{K}}_x) 
\end{align}
\begin{align}
    \dot{V}=&\frac{1}{2(\phi^{'2}_e -e^{\top}Pe)}\bigg(e^{\top}(PA_r+A_r^{\top}P)e+2e^{\top}P(-\frac{\dot{\phi}_e}{\phi_e}e \nonumber \\
    &+B\tilde{K}_xx+B\tilde{K}_rr +B\Delta u)\bigg) +\frac{\dot{\phi_e}e^{\top}Pe}{\phi_e(\phi^{'2}_e-e^{\top}Pe)} \nonumber \\
    & +tr(\tilde{K}_x^{\top}\Gamma_x^{-1}\dot{\hat{K}}_x) 
\end{align}
Employing the adaptive update law (\ref{updatelaws}),
\begin{align}
    \dot{V}\le -\frac{e^{\top}Qe}{2(\phi^{'^2}_e-e^{\top}Pe)}+ \frac{e^{\top}PB\Delta u}{\phi^{'^2}_e-e^{\top}Pe}
    \label{lyap11}
\end{align}
We now consider two cases depending on whether the controller is operating within the saturation limits (\textit{Case 1}) or is saturated (\textit{Case 2}). \\
\textit{Case 1.1:} $\|v(t)\|\leq \phi_u(t)$\\
    In this case, $u(t)=v(t)$ and $\Delta u(t)=0$, $\forall t \geq 0$.\\~\\
    \textit{Case 1.2:} $\|v(t)\|> \phi_u(t)$\\
    In this case, $u(t)=\phi_u(t)\displaystyle \frac{v(t)}{\|v(t)\|}$ and $\Delta u(t)=\frac{v(t)}{\|v\|}(\phi_u(t)-\|v(t)\|)$, which implies 
    \begin{align}
        \|\Delta u(t)\| & = \|v(t)\|-\phi_u(t)\\
                          & \leq
    \bar {K}_x(\|e\|+\|x_r\|)+ \bar{K}_r\bar{r}+\frac{|\dot{\phi}_e|}{\phi_e}\|B^{\dagger}\|\|e\|-{\phi}_u
    \label{ubound11}
    \end{align}
We will proceed with the Lyapunov analysis using Case $1.2$, as this case subsumes Case $1.1$, being the more general of the two.
Substituting (\ref{ubound11}) in (\ref{lyap11})
\begin{align}
    \dot{V}\le &-\frac{\lambda_{min}\{Q\}\|e\|^2}{2(\phi^{'^2}_e-e^{\top}Pe)}+\frac{\lambda_{max}\{P\}\|B\|\|e\|}{\phi^{'^2}_e-e^{\top}Pe}[\bar {K}_x\|e\|
    \nonumber \\
    &+\bar{K}_x{\mathcal{X}}_r+ \bar{K}_r\bar{r}-{\phi}_u] \\
    \leq & - \frac{\lambda_{min}\{Q\}\|e\|}{2(\phi^{'^2}_e-e^{\top}Pe)}\bigg[\frac{1}{\eta}\varrho -\sigma \|e\|\bigg]
    \label{lyapfunc2}
\end{align}
where, $\eta=\frac{\lambda_{min}\{Q\}}{2\lambda_{max}\{P\}\|B\|}$, $\varrho(t)=\phi_u(t)-\bar{K}_x{\mathcal{X}}_r(t)-\bar{K}_r\bar{r}$, $\sigma(t)=\sigma_0+\frac{|\dot{\phi}_e(t)|}{\eta\phi_e(t)}\|B^{\dagger}\|$ and  $\sigma_0=\frac{1}{\eta}(\bar{K}_x-\eta)$. Based on the sign of $\sigma_0$, we will now consider two cases.\\
\textit{Case 2.1 ($\sigma_0\geq0$):}
If the initial tracking error satisfies $\|e(0)\|<\phi_e(0)$, we have  $(\phi_e^{'^2}-e(0)^{\top}Pe(0))>0$ at $t=0$. To ensure that $\dot{V}(0)<0$, from the Lyapunov derivative in (\ref{lyapfunc2}), it is necessary that
\begin{align}
\phi_e(0) < \frac{\phi_u(0) - \bar{K}_x {\mathcal{X}}_r(0) - \bar{K}_r \bar{r}}{\bar{K}_x + \frac{|\dot{\phi_e}(0)|}{\phi_e(0)}\|B^\dagger\| - \eta}
\label{con1}
\end{align}
Recalling ${\phi}_e(t)={\phi}_x(t)-{\mathcal{X}}_r(t)$, the condition in (\ref{con1}) directly leads to feasibility condition C1. Moreover, when $\sigma_0=0$ , the controller gain satisfies $\bar{K}_x=\eta$, and under this scenario, at $t=0$, C1 reduces to
\begin{align}
    \phi_u(0)>|\dot{\phi}_e(0)|\|B^{\dagger}\|+\bar{K}_x{\mathcal{X}}_r(0)+\bar{K}_r\bar{r}
    \label{alphazero}
\end{align}
Note that, given the feasibility condition C1, $\varrho(t)>0$ $\forall t \ge 0$ is automatically guaranteed, which is a prerequisite for the feasibility of the controller under saturation. Employing C1 at $t=0$, (\ref{lyapfunc2}) satisfies
\begin{align}
    \dot{V}(0)< 0
    \label{uub}
\end{align}
\textit{Case 2.2 $(\sigma_0<0)$:} Following a similar argument as case 2.1, to achieve a stable result from (\ref{lyapfunc2}), we consider two scenarios. If $\varrho(t)\geq0$ we can prove (\ref{uub}) without any feasibility condition. On the other hand, if $\varrho(t)<0$, (\ref{uub}) can be proved from (\ref{lyapfunc2}) if the feasibility condition (\ref{C11}) holds.\\ 
Thus, for both cases we proved $\dot{V}(0)<0$. Now, define $t^\star:=\inf\{t>0:\; e^\top(t)Pe(t)=\phi_e^2(t)\}$. If $e(0)\in\Omega_e^{'}(0)$ and C1 holds,
$e(t)\in\Omega_e^{'}(t)$ for all $t\in[0,t^\star)$ and $\dot{V}(t)<0$ for all \(t\in[0,t^\star)\).
Hence \(V(t)\le V(0)<\infty\) on \([0,t^\star)\). Now, 
if $t^\star<\infty$, then as $t\rightarrow t^\star$ the BLF term $V_e(t)\rightarrow\infty$, and therefore $V(t)\to\infty$,
contradicting the bound \(V(t)\le V(0)<\infty\) on \([0,t^\star)\).
Therefore, \(t^\star=\infty\) and $e^{\top}(t)Pe(t)<\phi^{'2}_e(t)$ for all $t\ge 0$ (forward invariance).

For any $P>0$, $e^{\top}Pe\geq \lambda_{min}\{P\}\|e\|^2$ and it can be proved that $\|e(t)\|< \phi_e(t)$ $\forall t \geq 0$, i.e., the trajectory tracking error remains bounded within the pre-specified time-varying bound.\\
Further, since $x(t)=e(t)+x_r(t)$ and the reference model states and the trajectory tracking error is bounded, i.e. $\|x_r(t)\|\leq {\mathcal{X}}_r(t)$, $\|e(t)\|<\phi_e(t)$, it can be easily shown that the proposed controller guarantees the plants states to be bounded within the user defined safe set
\begin{align}
    \|x(t)\|< \phi_e(t)+{\mathcal{X}}_r(t)< \phi_{x}(t) && \forall t \geq 0
\end{align}
Since the closed loop tracking error as well as the controller parameter estimation errors remain bounded and $K_x$ and $K_r$ are constants, it can be concluded that the estimated parameter is also bounded i.e. $\hat{K}_x(t)\in \mathcal{L}_{\infty}$ followed by ensuring the plant state $x(t)$ and control input $u(t)$ to be bounded for all time instances. Thus, the proposed controller guarantees that all the closed-loop signals are bounded. We have shown that the TVBLF remains finite, which in turn implies that the state trajectory never reaches the constraint boundary. Consequently, all system signals remain bounded, and both constraints are satisfied for all time.
\end{proof}
\begin{remark}[Existence and Uniqueness of Solutions]
The closed-loop dynamics under the projection operator is piecewise locally Lipschitz in $(x,\hat{K}_x)$, ensuring the existence and uniqueness of a maximal solution on $[0,\tau_{\max})$ \cite[Theorem~54]{Sontag}, ). The Lyapunov analysis in Theorem~\ref{maintheorem} further guarantees that the trajectory remains in a compact, forward-invariant subset of the admissible domain, thereby ruling out finite escape time. Hence, $\tau_{max}=\infty$ and the solution exists and is unique for all $t\geq 0$.
\end{remark}
\begin{corollary}
Consider the MIMO LTI plant with unknown unmatched disturbance.
\begin{align}
    \dot{x}=Ax+Bu+d
    \label{plant_disturbance}
\end{align}
where the disturbance $d(t)\in\mathbb{R}^n$ is bounded such that $\sup_{t\ge0}\|d(t)\|<\bar{d}$ and $\bar{d}>0$ is a known constant. Under the projection-based adaptive law \eqref{updatelaws},
the time-varying state and input constraints \eqref{statecon}--\eqref{incon} remain satisfied
and all closed-loop signals are bounded for all $t\ge0$ provided the following
modified feasibility condition holds for all $t\ge0$:
\begin{equation}
\label{C1d}
{\phi}_u(t)>{\phi}_x(t) (\bar{K}_x-\eta)+ |\dot{\phi}_e(t)|\|B^{\dagger}\|+\eta \mathcal{X}_r(t) +\bar{K}_r \bar{r} +\frac{\bar{d}}{\|B\|}
\end{equation}
\end{corollary}
\begin{proof}
   The result follows by incorporating the disturbance term into the Lyapunov
analysis of Theorem~\ref{maintheorem}. Detailed proof is omitted for brevity.
\end{proof}


\section{Discussion on the Feasibility Condition (C1)} \label{feas}
To ensure feasibility under time-varying constraints, we derive a sufficient condition (C1) that guarantees the existence of a feasible control policy. This condition imposes a lower bound on the input constraint that has to be satisfied for all time instances and establishes an explicit trade-off between the state constraint $\phi_x(t)$, the input constraint $\phi_u(t)$, and the rate of change of the error bound $|\dot{\phi}_e(t)|$. Unlike prior works requiring real-time feasibility monitoring \cite{worthmann2020funnel} or optimization routines, this approach offers a computationally efficient offline verifiable feasibility guarantee. From the Lyapunov stability analysis, the feasibility condition (C1) takes the form
    \begin{align}
{\phi}_u(t)>\alpha_1{\phi}_e(t)+ \alpha_2|\dot{\phi}_e(t)|+\beta \label{f_con1}
\end{align}
where 
\begin{align}
    &\alpha_1 =\bar{K}_x-\frac{\lambda_{min}\{Q\}}{2\lambda_{max}\{P\}\|B\|} \label{alpha1}\\
    &\alpha_2= \|B^{\dagger}\| \label{alpha2}\\
    & \beta = \frac{\bar{\mathcal{X}_r}\lambda_{min}\{Q\}}{2\lambda_{max}\{P\}\|B\|}+\bar{K}_r\bar{r} \label{beta}
\end{align}
the feasibility condition
splits the input budget into three parts:
(i) error envelope  $\alpha_1 \phi_e$,
(ii) rate of change of the error envelope  $\alpha_2 |\dot{\phi}_e|$,
and (iii) reference/model offset $\beta$.

\subsection{Significance of the Coefficients ($\alpha_1$, $\alpha_2$ and $\beta$)}
The coefficients in \eqref{f_con1} quantify three independent mechanisms.
\begin{enumerate}
    \item [(i)] The sign of $\alpha_1$ implies the closeness between the system matrix \(A\) of the plant and \(A_r\) of the reference model, in the two-norm sense.
An implication of $\alpha_1>0$ (or, $\sigma_0>0$), when combined with the matching condition (\ref{mc11}), is the following
\begin{align}
    \|A_r - A\| > \frac{\lambda_{\text{min}}\{Q\}}{2 \lambda_{\text{max}}\{P\}\|B\|}
    \label{eq123}
\end{align}
which indicates that the distance between \(A\) and \(A_r\) in the two-norm sense is relatively large. In this case (Case 2.1), for a larger error constraint at any particular time instant, the lower bound of the input constraint increases. In contrast, for $\alpha_1<0$ (or, $\sigma_0<0$), the inequality in (\ref{eq123}) is reversed, which implies that the distance between \(A\) and \(A_r\) in the two-norm sense is relatively less. For this case (Case 2.2), the lower bound on the input constraint decreases if we increase the error constraint. Furthermore, if $\alpha_1=0$, the condition C1 simplifies to \eqref{alphazero}, which is independent of $\phi_e(t)$, but still depends on $|\dot{\phi}_e(t)|$ and $\beta$.
\item [(ii)] $\alpha_2$ measures how `responsive' the plant is to the input. Since $\|B^{\dagger}\|=\frac{1}{\sigma_{min}\{B\}}$, for a larger $\sigma_{min}\{B\}$\footnote{\(\sigma_{\min}(B)\) is the smallest singular value of the input matrix \(B\).
} yields a smaller $\alpha_2$. In this case, fast variations in the state bound $|\dot{\phi}_e(t)|$ require only a modest increase in the input budget, making feasibility easier to achieve; whereas a smaller $\sigma_{min}\{B\}$ increases $\alpha_2$ and raises the input demand to satisfy the feasibility condition.
\item [(iii)] $\beta$ aggregates the effect of bounded reference signals and their gains; larger reference magnitudes increase the constant offset in the input budget.
\end{enumerate}

\subsection{Steady-state and Transient Limits}
The rate of change in the error constraint ($|\dot{\phi}_e(t)|$) quantifies how rapid variations in the error constraint impact the required input magnitude. A faster changing error bound requires a proportionally larger input and vice-versa. When $\dot{\phi}_e(t)\to0$ and $\phi_e(t)\to\phi_e^\infty$, C1 reduces to
\begin{equation}
\phi_u^\infty > \alpha_1 \phi_e^\infty + \beta
\label{steadystate}
\end{equation}
where $\phi^{\infty}_e>0$ and $\phi^{\infty}_u>0$ are the user-defined steady-state bound of the error and input, respectively. Considering $\dot{\phi}_e\equiv0$ recovers MRAC with fixed state and input constraints as a special case of our framework, with C1 reducing to the steady-state feasibility condition \eqref{steadystate} and $\sup_{t\ge0}\|u(t)\|\leq\phi_u^{\infty}$.

\subsection{Conservatism of C1}
C1 is a sufficient offline feasibility certificate. Unlike PPC or FC, which typically channel saturation deficiency by modifying the reference model or reshaping the boundary online, our approach accommodates it in the prescribed input budget. The certificate is derived under worst-case assumptions, including the maximum admissible disturbance \eqref{C1d} or the largest possible saturation deficiency. Consequently, at instants when the actual disturbance or the deficiency is small, C1 may require demand input authority or a looser error envelope than is actually needed. This is a trade-off for an offline verifiable condition that preserves the given envelops and reference trajectory while ensuring forward-invariance. 
\subsection{Only State Constraint} 
If no constraint is imposed on the control input, actuator saturation is inactive and this scenario represents Case~1.1, where $\Delta u(t)=0$. If the initial error satisfies $e(0)\in \Omega_e(0)$, feasibility of the control policy follows directly and no additional condition is required.
    \subsection{Only Input Constraint}
    If no state constraint is considered, ${\phi_e}(t)=\dot{\phi}_e(t)=0$ $\forall t \ge 0$.
Instead of TVBLF-based adaptive update laws \eqref{updatelaws}, we can employ the classical adaptive laws \eqref{MRAC} to achieve the control objective. The feasibility condition C1 reduces to 
\begin{align}
\phi_u(t)>\eta{\mathcal{X}}_r(t)+\bar{K}_r\bar{r}
\end{align}

\section{Designing Constraints as Performance Functions}
\label{splcases}
\label{sec5}
The proposed framework subsumes both PPC and FC as special cases under specific choice of performance functions. To impose a pre-defined transient and steady-state behavior on the tracking error (consequently on the plant state) and the control input, we introduce generalized PPF, \( \phi_{e,u}(t): [0, \infty) \to (\phi^{\infty}_{e,u}, \phi^{0}_{e,u}] \), where \( \phi^{0}_{e,u} > 0 \) and ${\phi}^{\infty}_{e,u}>0$ denote the initial and final bounds, respectively and $\phi^{0}_{e,u}>\phi^{\infty}_{e,u}$. The performance function is defined as
\begin{equation}
    \phi_{e,u}(t) := (\phi^{0}_{e,u} - \phi^{\infty}_{e,u} )\left(1 + \kappa_{e,u} t^{\nu_{e,u}}\right)^{-1} + \phi^{\infty}_{e,u}
    \label{ppf}
\end{equation}
where \( \kappa_{e,u}, \nu_{e,u} \in \mathbb{R} \) are positive design parameters that govern the rate and shape of the performance envelope, respectively. The parameter \( \kappa_{e,u} \) controls the rate of decay of the performance function, directly influencing how quickly the error approaches the desired bound. A higher value of \( \kappa_{e,u} \) results in faster initial convergence, while a smaller value slows down the decay rate. On the other hand, the parameter \( \nu_{e,u} \) shapes the curvature of the convergence profile. For values of \( \nu_{e,u} < 1 \), the performance envelope exhibits a concave profile, characterized by a rapid initial decrease followed by a slower decay toward the steady-state limit. In contrast, when \( \nu_{e,u} > 1 \), the envelope becomes convex, leading to a slower initial reduction but a sharper drop near the steady state. The case \( \nu_{e,u} = 1 \) corresponds to an exponential decay. \\
The convergence time, at which the performance function reaches a specified error threshold \( \epsilon_{e,u} \), where \( \epsilon_{e,u} > \phi_{e,u}^{\infty} \), can be obtained by solving \( \phi(t)_{e,u} = \epsilon_{e,u} \). Substituting into the performance function expression yields
\begin{equation}
    \epsilon_{e,u} = (\phi^{0}_{e,u} - \phi^{\infty}_{e,u}) \left(1 + \kappa_{e,u} t^{\nu_{e,u}}\right)^{-1} + \phi^{\infty}_{e,u}
    \label{ppf1}
\end{equation}
Solving (\ref{ppf1}), we get the convergence time 
\begin{equation}
    \tau_{_{e,u}} = \left[ \frac{1}{\kappa_{e,u}} \left( \frac{\phi^{0}_{e,u} - \phi^{\infty}_{e,u}}{\epsilon_{e,u} - \phi^{\infty}_{e,u}} - 1 \right) \right]^{\frac{1}{\nu_{e,u}}}.
\end{equation}
For a fixed \( \tau \), increasing \( \nu \) reduces \( \kappa \), whereas decreasing \( \nu \) requires a larger \( \kappa \) to maintain the same convergence time.

\begin{figure}[H]
    \centering
    \subfigure[]{
        \includegraphics[width=0.42\linewidth]{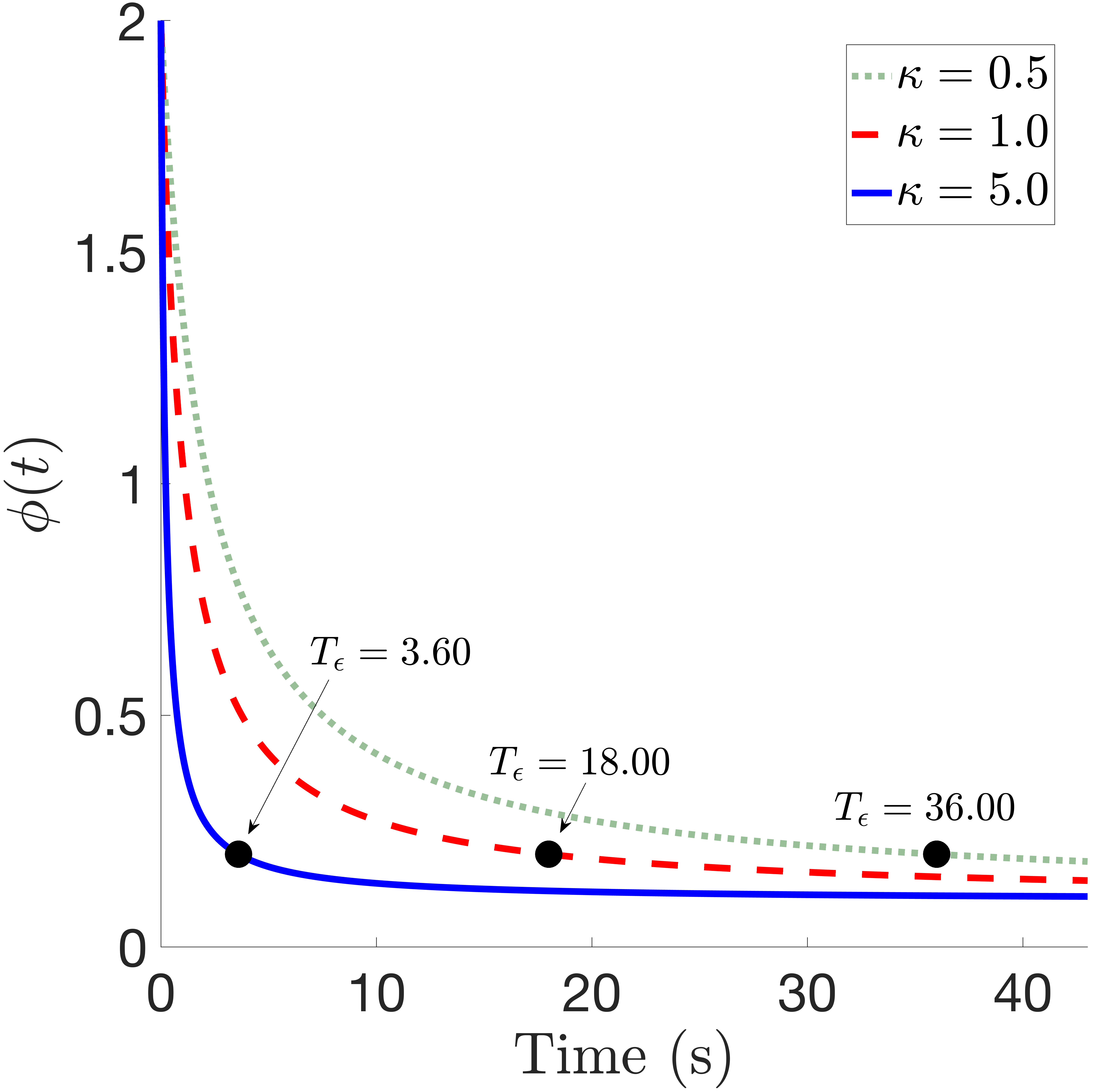}
    }
    \subfigure[]{
        \includegraphics[width=0.42\linewidth]{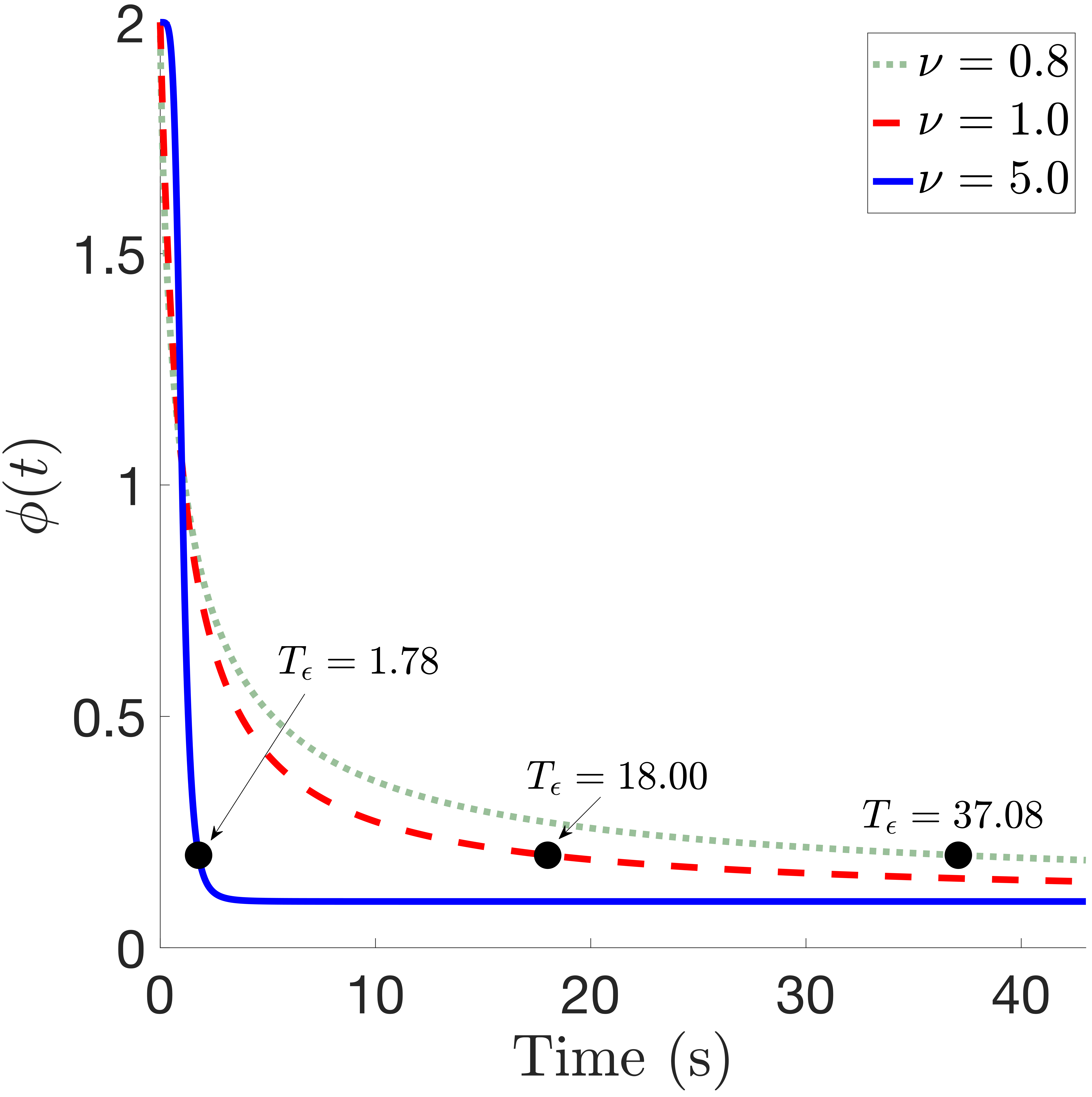}
    }
    \caption{Effect of varying (a) $\kappa$  and (b) $\nu$ on the performance function $\phi_{e,u}(t)$, where $\phi_{e,u}^{0}=2$, $\phi_{e,u}^{\infty}=0.1$ and  $\epsilon=0.2$. $\nu=1$ and $\kappa=1$ are fixed in (a) and (b), respectively.}
\end{figure}
\begin{remark}
    While standard PPC or FC-based formulations enforce time–varying error bounds, input limits are treated only indirectly, i.e., by online reshaping the funnel \cite{fc3}, or by modifying the reference trajectory \cite{ppc3}. In contrast, our approach co–designs time–varying state and input envelopes and comes with an offline, verifiable feasibility condition that certifies a feasible pair $(\phi_x(t),\phi_u(t))$ for the given plant and reference model.
\end{remark}

\section{Simulation Results}
To demonstrate the efficacy of the proposed algorithm, we consider two examples. For both examples, we select the time-varying error (transformed from state constraint) and input constraints as PPFs, as discussed in Section \ref{sec5}. Given the plant and the reference model, to assess the feasibility of the error (transformed from state constraint) and input constraint pairs, we may adopt the following approaches:
\begin{enumerate}
    \item For any given $\phi_e(t)$ and $\phi_u(t)$, evaluate feasibility directly using feasibility condition C1.
    \item For any given $\phi_e(t)$, compute the minimum value of $\phi_u(t)$ that satisfies C1, and vice-versa.
    \item If a given $\{\phi_e(t),\phi_u(t)\}$ pair is infeasible but the user can allow relaxation of one or both variables, assign appropriate weights to $\gamma$ such that C1 is satisfied. 
\end{enumerate}
For simulation, we adopt the first approach.
\subsection{Example 1}
\label{example1}
We consider the approximated lateral dynamics of a linear bicycle model \cite{rajamani2012vehicle} at low-forward speed, neglecting roll-yaw coupling, given by
\begin{align}
    &\dot{x}_1=x_2\nonumber\\
    &\dot{x}_2=kx_1+cx_2+u
    \label{cycle}
\end{align}
where $x_1(t)$ represents the lateral displacement, $x_2(t)$ is the lateral velocity and  $x(t)=\begin{bmatrix}
    x_1(t) & x_2(t)
\end{bmatrix}^{\top}\in\mathbb
{R}^{2}$. The control input is denoted by $u(t)\in \mathbb{R}$ and $k,c\in \mathbb{R}_{>0}$ are effective lateral stiffness ( represents restoring force per displacement) and damping (depends on speed and yaw) coefficients, respectively. At low speed, $k$ and $c$ are typically positive, which can result in an unstable open-loop system. A virtual desired stable reference model is chosen which follows
\begin{align}
   &\dot{x}_{r_1}=x_{r_2}\nonumber\\
    &\dot{x}_{r_2}=k_rx_{r_1}+c_rx_{r_2}+r
    \label{cycleref} 
\end{align}
where, $x_r(t)=\begin{bmatrix}
    x_{r_1}(t) & x_{r_2}(t)
\end{bmatrix}^{\top}\in\mathbb
{R}^{2}$ represents the reference model state, $r(t)\in \mathbb{R}$ denotes the reference input and $k_r,c_r\in \mathbb{R}$ are the stiffness and damping coefficients, respectively.\\
The time-varying error and input constraints are given as
\begin{align}
    &\phi_e(t)=(\phi_e^0-\phi_e^{\infty})(1+\kappa_e t^{\nu_e})^{-1}+\phi_e^{\infty} \label{phie1}\\
    &\phi_u(t)=(\phi_u^0-\phi_u^{\infty})(1+\kappa_u t^{\nu_u})^{-1}+\phi_u^{\infty}
    \label{phiu1}
\end{align}
where $\phi^0_{e,u},\phi_{e,u}^{\infty},\kappa_{e,u},\nu_{e,u}$ are defined in Section \ref{sec5}. For simulation, we have considered $\phi_e^0=0.8,\:  \phi_e^{\infty}=0.05,\:\kappa_e=1.2,\: \nu_e=1$,\: $\phi_u^0=5,\:\phi^{\infty}_u=1.7,\: \kappa_u=1,\: \nu_u=1$, $k=2 $ N/m, $c= 1.5$ Ns/m, $k_r=-1$ N/m $c_r=-2$ Ns/m, $r(t)=\sin(0.5t)$ and  $\Gamma_x=5$. We verify that the feasibility condition C1 is satisfied.\\
Fig. $4$a shows that the proposed method ensures that the plant state tracks the reference trajectory while satisfying the time-varying constraint. Further, the control input also remains within the pre-specified time-varying bound (Fig. $4$e).
\begin{figure*}[h!]
    \centering
    \vspace{0.2em}
    \subfigure[Tracking performance]{\includegraphics[width=0.3\linewidth]{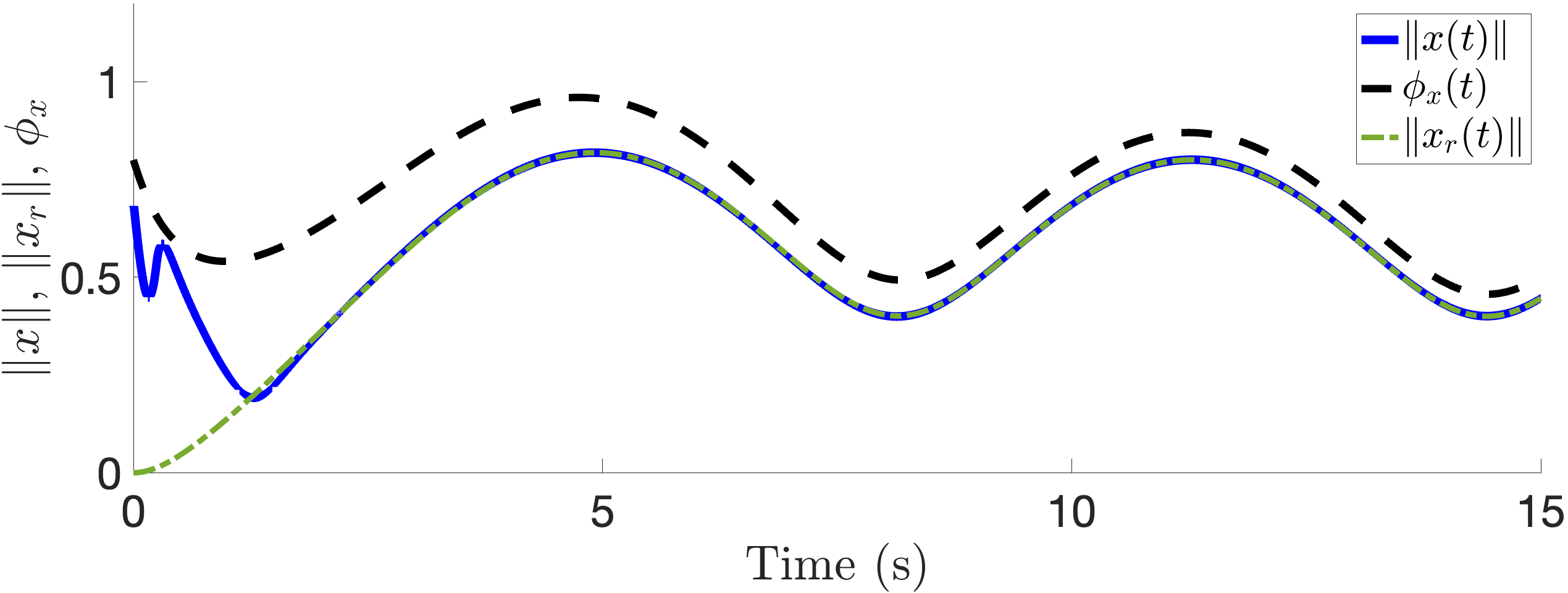}}
    \subfigure[Tracking error]{\includegraphics[width=0.3\linewidth]{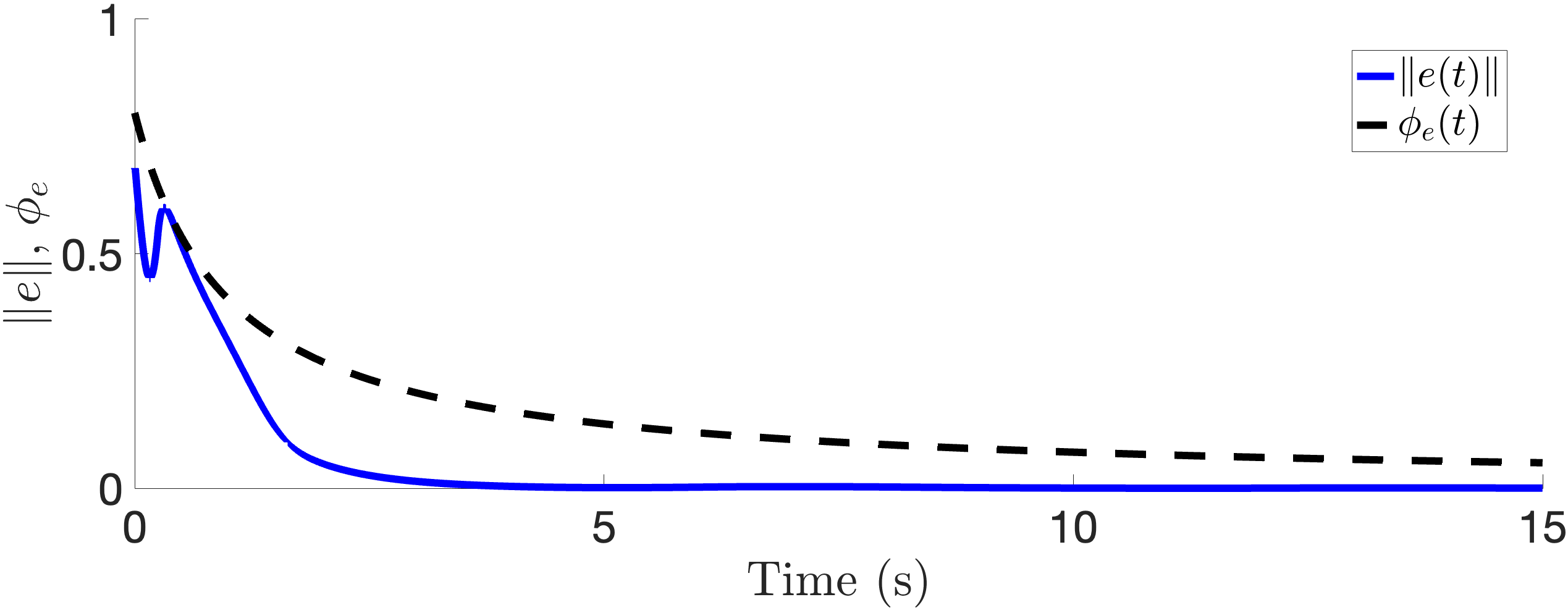}}
    \subfigure[Control input]{\includegraphics[width=0.3\linewidth]{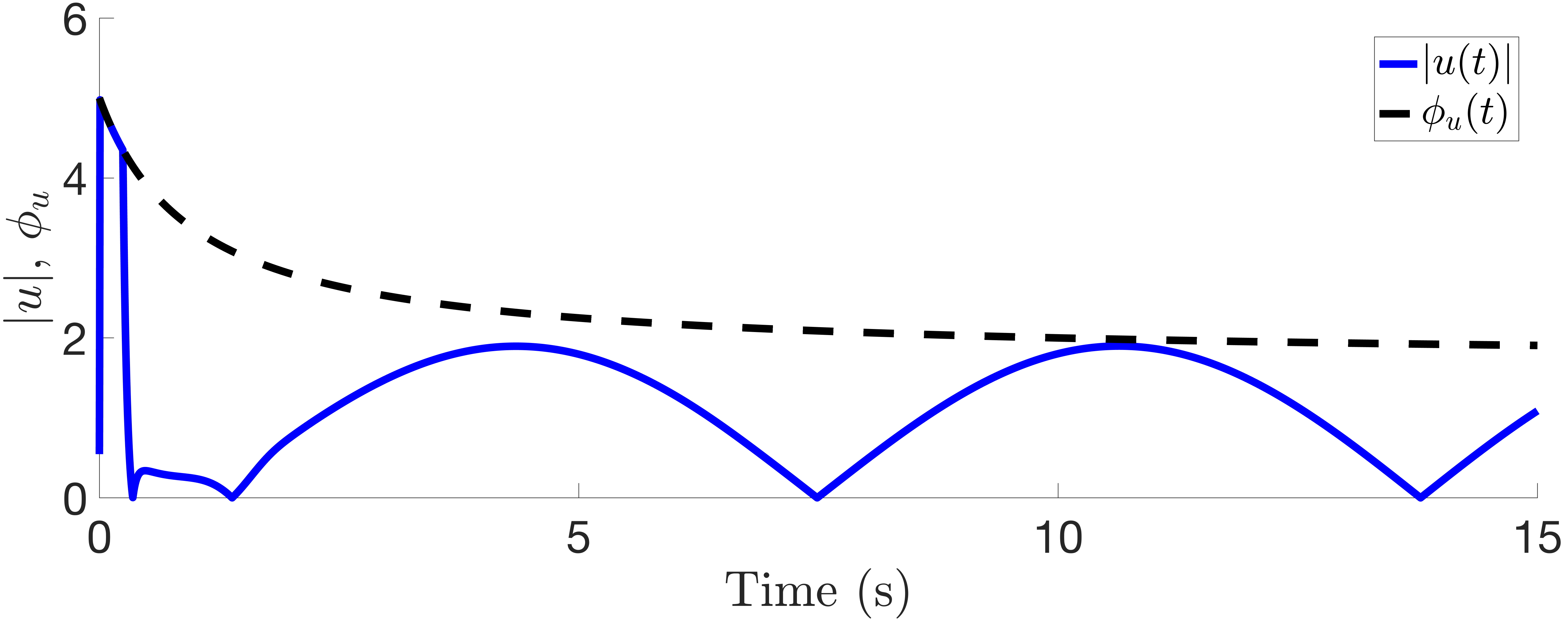}}
    \subfigure[Tracking performance]{\includegraphics[width=0.3\linewidth]{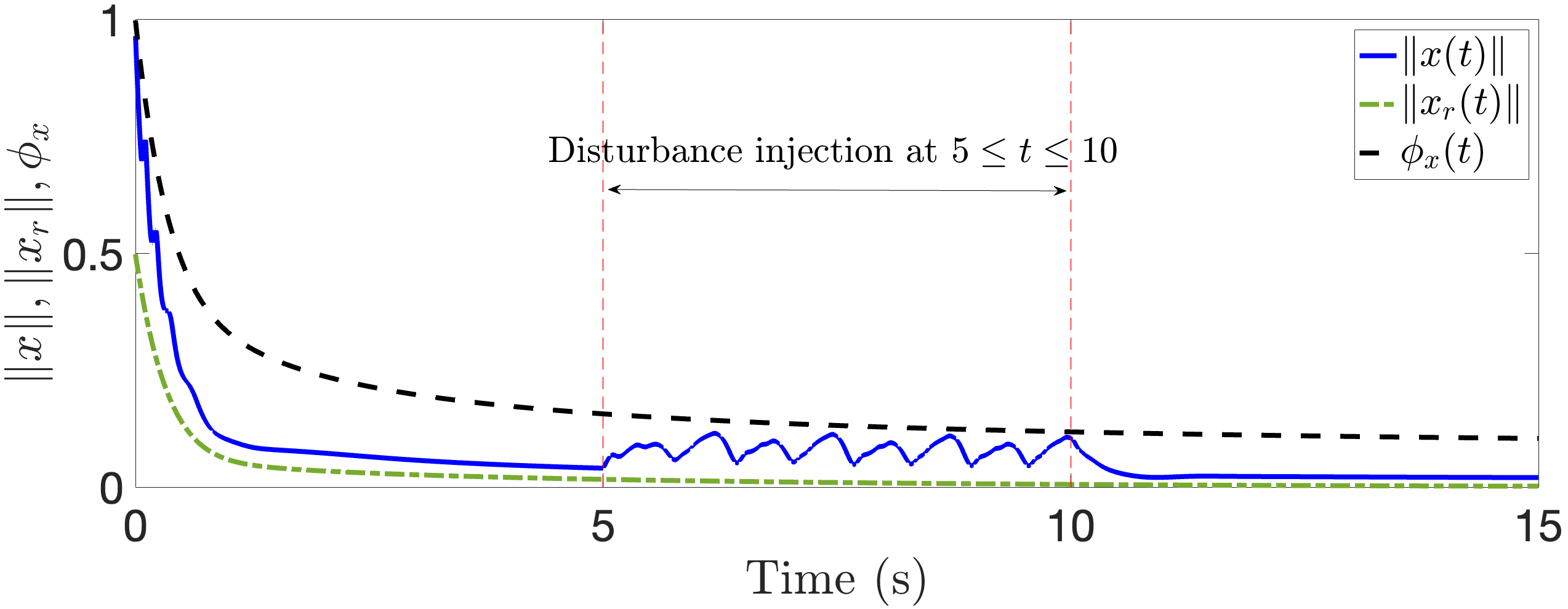}}
    \subfigure[Tracking error]{\includegraphics[width=0.3\linewidth]{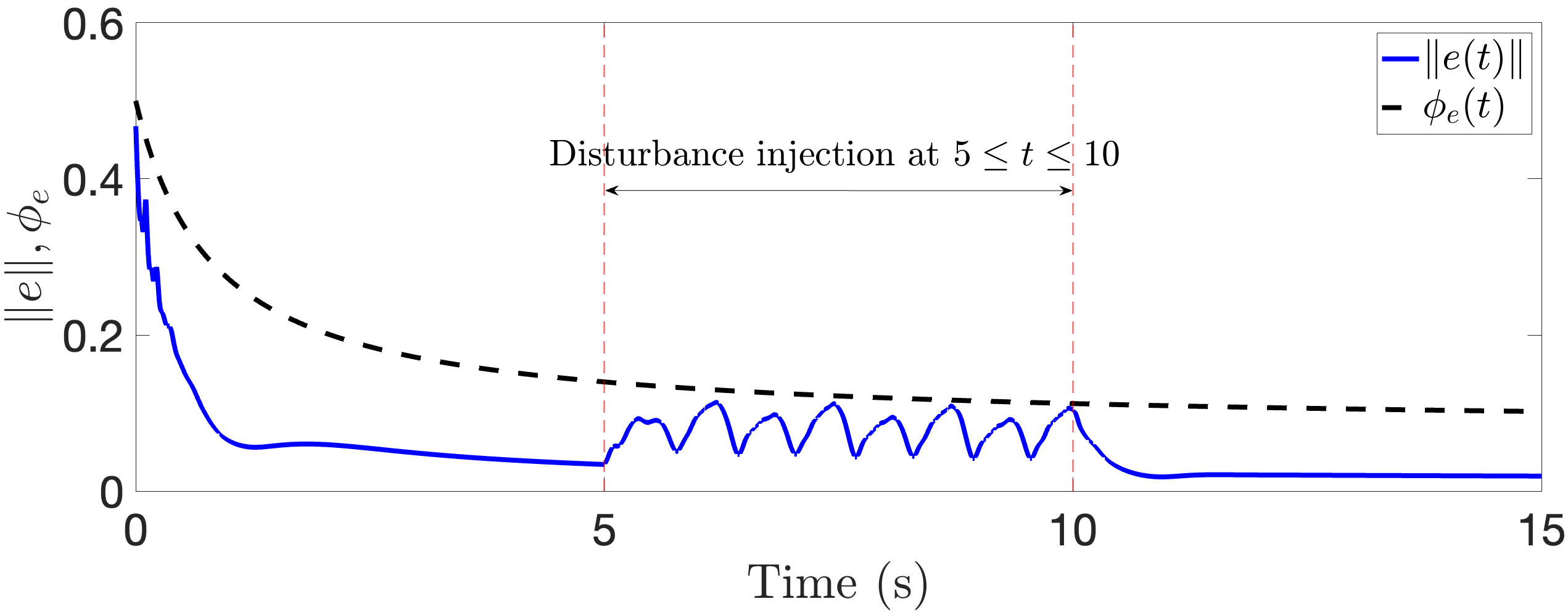}}
    \subfigure[Control input]{\includegraphics[width=0.3\linewidth]{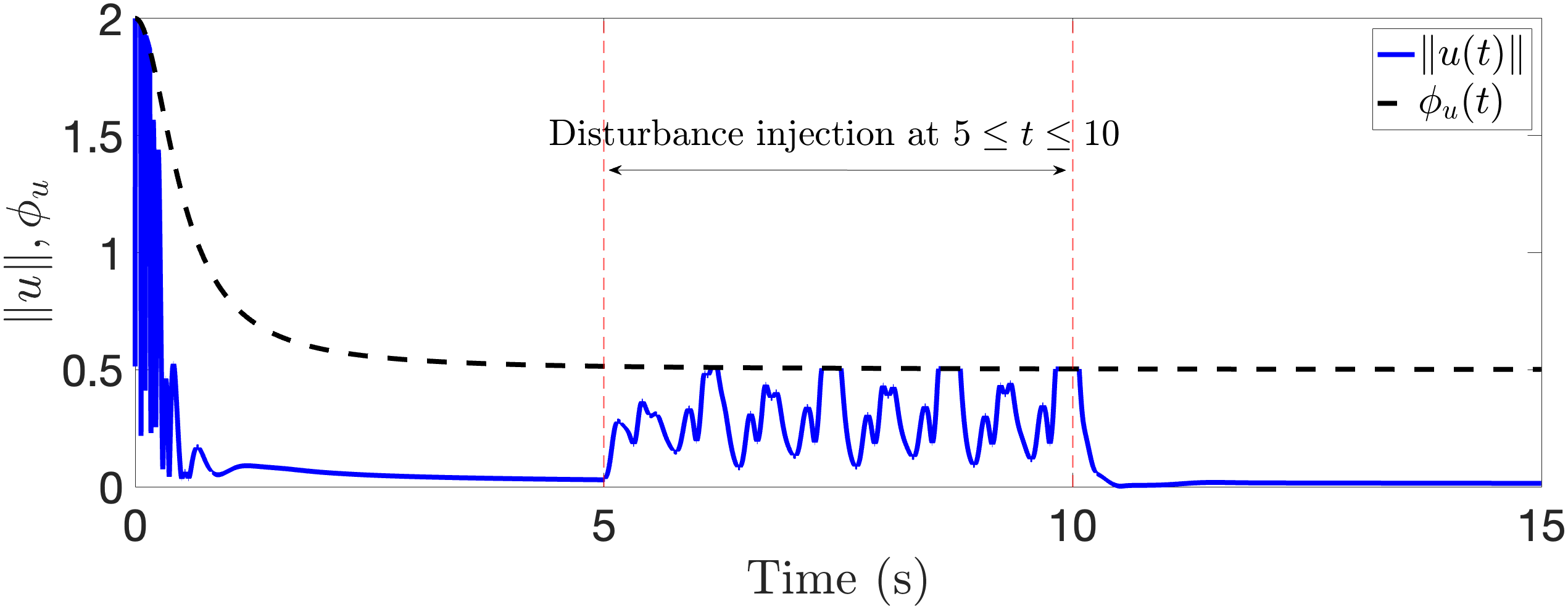}}
     \subfigure[Tracking performance]{\includegraphics[width=0.3\linewidth]{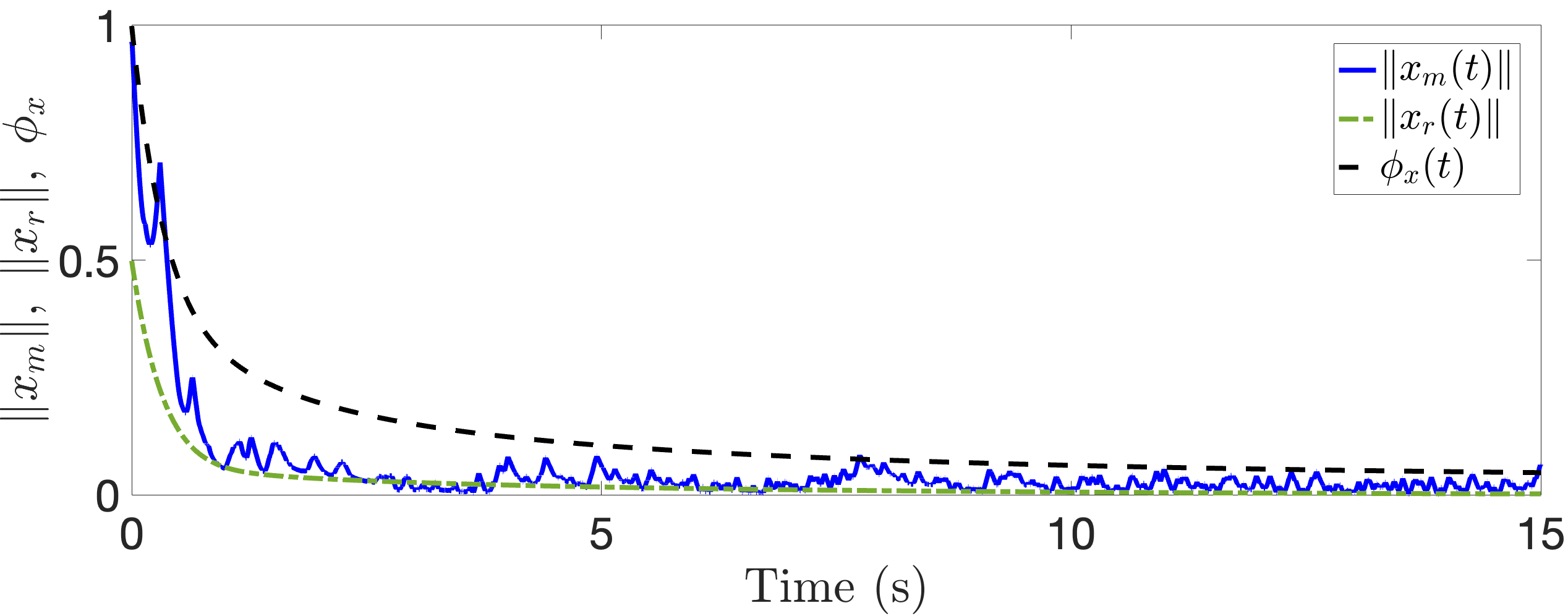}}
     \subfigure[Tracking error]{\includegraphics[width=0.3\linewidth]{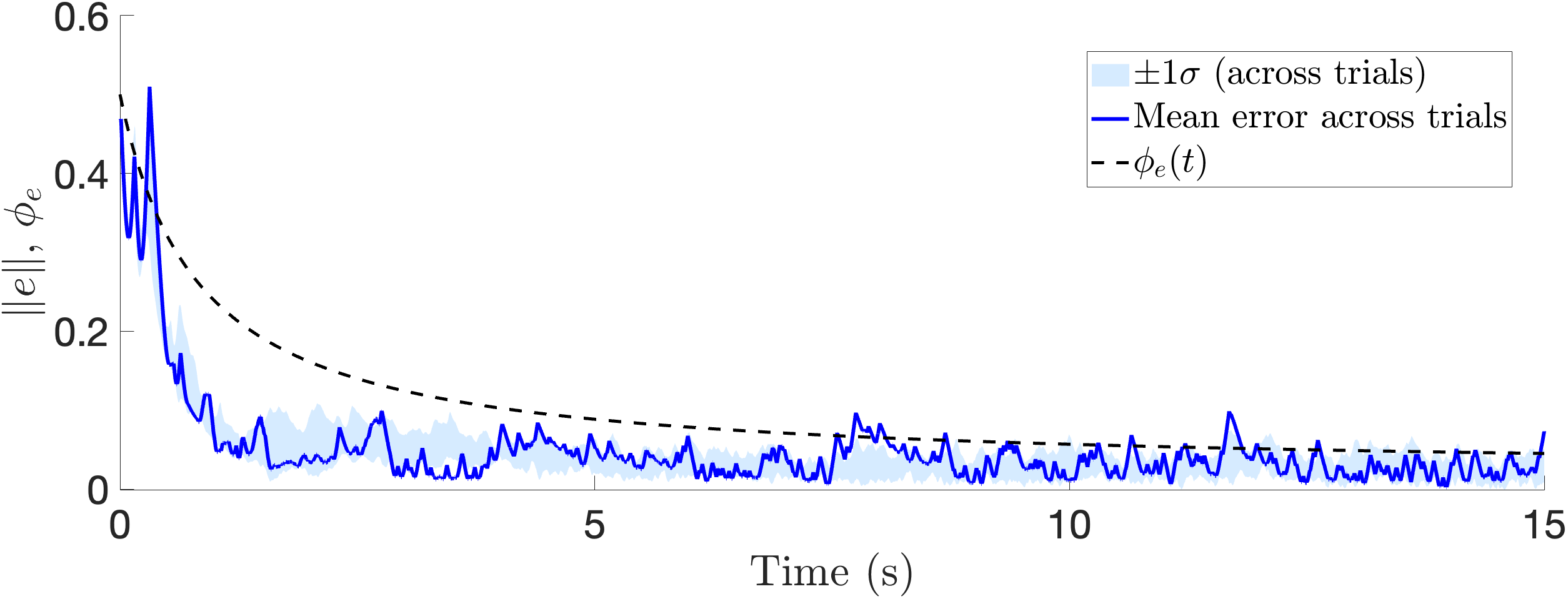}}
     \subfigure[Control input]{\includegraphics[width=0.3\linewidth]{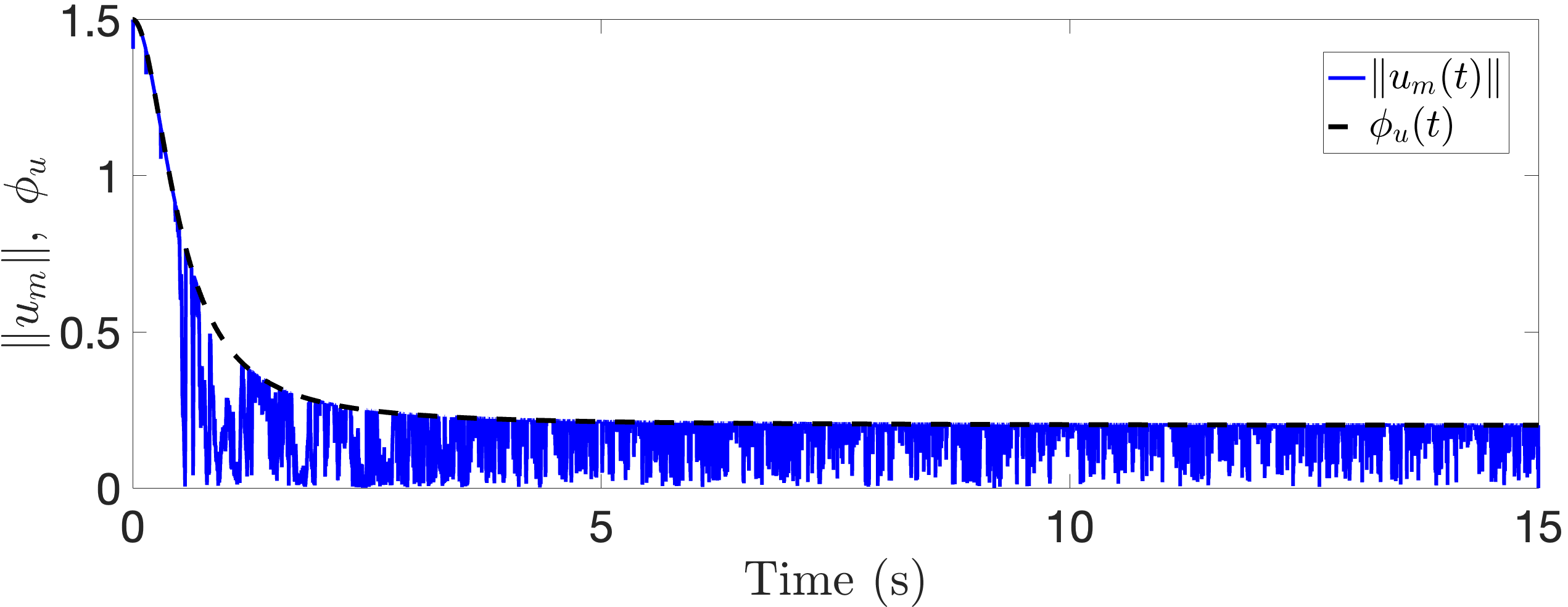}}
    \caption{Example 1 ((a)-(c)): performance using the proposed method; Example 2 ((d)-(f)): proposed method in presence of disturbance \eqref{disturbance}; Example 2 ((d)-(f)): proposed method in presence of disturbance \eqref{disturbance}. Example 2 ((g)-(i)): proposed method in presence of measurement noise \eqref{state_noise} with $\sigma^2=0.05$. $u_m(t)\in\mathbb{R}^m$ refers to the corresponding input signal with measurement noise.
    }
    \label{nosat_state}
\end{figure*}

\subsection{Example 2}
\label{example2}
We consider an unstable MIMO LTI plant and a stable reference model that follow the dynamics of \eqref{plant_disturbance} and \eqref{ref}, respectively, where
\begin{align}
    A=&\begin{bmatrix}
        -3 & 1.5 &0 &0\\
        0 &-3 &1.5 &0\\
        0 &0 &-3 &-1.5\\
        0 &0 &0 &3
    \end{bmatrix} &&
    B=\begin{bmatrix}
        0 &0\\
        1 &0\\
        0 &0\\
        3 &1
    \end{bmatrix}\\
    A_r=&\begin{bmatrix}
        -2 &-1 &0 &0\\ 
        0 &-2 &-1 &0\\
        0 &0 &-2 &-1\\
        0 &0 &0 &-2
    \end{bmatrix} &&
    B_r=\begin{bmatrix}
        0 &0\\
        1 &0\\
        0 &0\\
        1 &1
    \end{bmatrix}
\end{align}
We impose time-varying state \eqref{statecon} and input constraints \eqref{incon} where $\phi_e(t)$ and $\phi_u(t)$ are designed as \eqref{phie1} and \eqref{phiu1}, respectively, and $\phi_e^0=0.5$, $\phi_e^{\infty}=0.08$, $\kappa_e=1.2$, $ \nu_e=1$, $\phi_u^0=2$, $\phi^{\infty}_u=0.5$ $\kappa_u=4$, $\nu_u=2$. The other simulation parameters are chosen as,  $r(t)=0.02[\exp(-0.2t);\exp(-t)]$,$Q=0.1\mathbb{I}_4$, $\bar{K}_x=10$, $\bar{K}_r=2.5$ and  $\Gamma_x=\begin{bmatrix}
    0 &2\\
    2 &0
\end{bmatrix}$.
\subsubsection{Performance under Bounded External Disturbance}
To evaluate the robustness of the proposed method against external disturbances, 
we inject the following bounded disturbance during a finite time window:

\begin{equation}
  d(t)=d_1(t)\,\mathbf{1}\{5\le t\le 10\}
  \label{disturbance}
\end{equation}
where $d_1(t)=0.5[
            \sin(10t);
            \cos(10t);
            \sin(5t);
            2]\in \mathbb{R}^4$ is  bounded disturbance. We verify that the feasibility condition \eqref{C1d} is satisfied.

As shown in Fig.~$5$a-$5$b, the plant state and the corresponding trajectory tracking error remain within a dynamically shrinking envelope, even in presence of bounded external disturbance \eqref{disturbance}. Further, the proposed controller ensures that the required control input is also confined within the user-defined time-varying constraint, as shown in Fig. $5$c. 
\subsubsection{Performance under Measurement Noise}\label{subsec:meas-noise}
\begin{figure}[!t]
\centering

\begin{minipage}[t]{0.38\linewidth}
\vspace{0pt}
\centering
\renewcommand{\arraystretch}{1.15}
\begin{tabular}{cc}
\toprule
$\sigma^2$ & ${P}_{\text{avg}}$ \\
\midrule
0.001 & 0.99 \\
0.01 & 0.95 \\
0.05 & 0.86 \\
0.08 & 0.75 \\
0.1 & 0.68 \\
\bottomrule
\end{tabular}

\vspace{1ex}
\captionof{table}{}
\label{table_noise}
\end{minipage}
\hfill
\begin{minipage}[t]{0.58\linewidth}
\vspace{0pt}
\centering
\includegraphics[width=\linewidth]{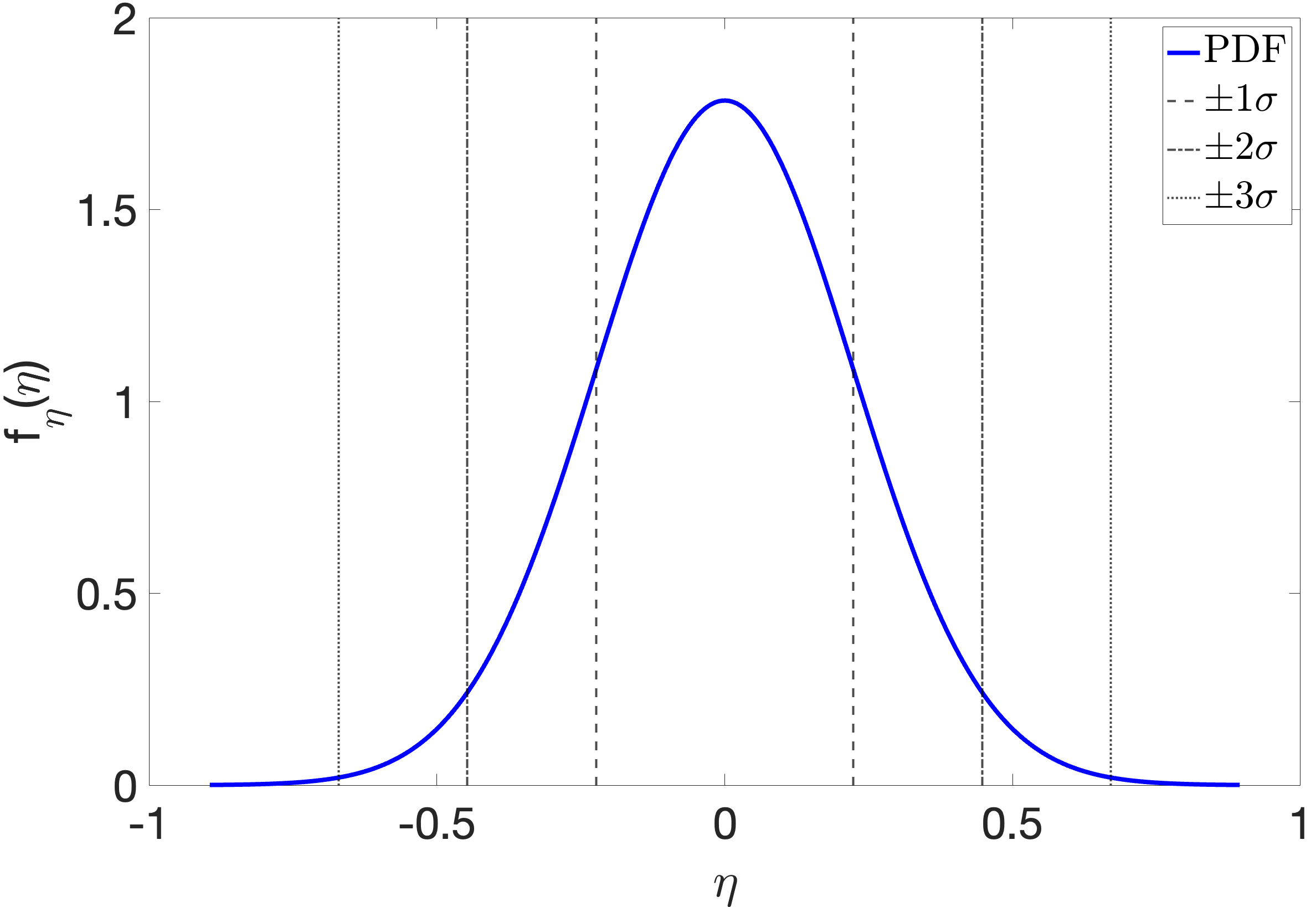}

\captionof{figure}{}
\label{noise_curve}
\end{minipage}

\vspace{1ex}
\caption*{Table~\ref{table_noise}: Effect of $\sigma^2$ on ${P}_{\text{avg}}$; Fig.~\ref{noise_curve}: Probability distribution function (PDF) of the measurement noise with $\sigma^2=0.05$.}
\label{fig:noise_table_plot_combo}
\end{figure}

To assess robustness to measurement noise, we inject zero-mean white Gaussian noise in the measurement channel. The plant evolves with the true state $x(t)$, while the controller and adaptation laws use the noisy state measurement given by
\begin{equation}
    x_m(t_k) = x(t_k) + \eta_k,\qquad
    \eta_k{\sim} \mathcal{N}\!\big(0,\sigma^2 I_n\big)
    \label{state_noise}
\end{equation}
where  $t_k$ represents the $k$-th time instant, $\eta_k$ is the independent and identically distributed noise at sample $k$ and $\sigma^2$ denotes the variance. We evaluate constraint satisfaction on the measured error $e_m(t_k)=x_m(t_k)-x_r(t_k)$ via the constraint violation margin
\begin{equation}
    h_m(t_k)=\phi_e^{'2}(t_k)\;-e_m^\top(t_k)Pe_m(t_k).
\end{equation}
For a fixed $\sigma^2$ and a time window $\mathcal{K}=\{k:t_k\in[t_a,t_b]\}$, the empirical time-averaged probability of constraint satisfaction of the measured error is estimated with Monte-Carlo simulation ($N=1000$) as
\begin{equation}
P_{avg}=
    \frac{1}{N|\mathcal{K}|}
    \sum_{i=1}^{N}
    \sum_{k\in\mathcal{K}} \mathbf{1}\!\left\{h_m^{(i)}(t_k) > 0\right\}
\end{equation}
where $N$ denotes the number of Monte Carlo trials. Each trial corresponds to one independent simulation with a different realization of the measurement noise sequence, and averaging across $N$ trials yields an empirical estimate of the satisfaction probability.\\
Since $d(t)\equiv0$, we tighten the input constraint by setting $\phi^0_u=1.5$, $\phi_u^{\infty}=0.2$ and verify that C1 is satisfied. Although measurement noise is not included in the analysis (and thus not accounted for in the feasibility condition), Monte-Carlo simulations with 
$\sigma^2=0.05$ show that the measured state/error remains within the time-varying bound for approximately $86\%$ of the evaluation window (Fig. $4$(g)–$4$(h), Table~\ref{table_noise}), while the time-varying input constraint is respected for all time (Fig. $4$(i)).

\begin{remark}
Since $e_m=e+\eta$, we have $\mathbb E[e_m^\top P e_m]=e^\top P e+\sigma^2\mathrm{tr}(P)$, so the expected margin $\mathbb E[h_m]$ shrinks linearly with $\sigma^2$. 
Moreover, for a larger $\sigma^2$, $\mathrm{Var}(e_m^\top P e_m)$ increases leading to a smaller $P_{avg}$, as shown in Table~\ref{table_noise}.
\end{remark}

\section{Conclusion}
In this paper, we have designed a  novel MRAC architecture for multivariable LTI systems by integrating time-varying BLF with a saturated controller which guarantees both the plant state and the control input remain bounded within user-defined time-varying safe sets. The development of a verifiable, offline feasibility condition provides a practical and computationally efficient tool for constrained control design, establishing a  relationship between state/input limitations and controller feasibility. We validate the proposed approach through simulation, comparing it with several existing related methods. Future research will focus on extending this powerful framework to broader classes of uncertain nonlinear systems.

\bibliographystyle{ieeetr}
\bibliography{ref}

\begin{thebibliography}{10}

\bibitem{mrac3}
S.~Sastry, M.~Bodson, and J.~F. Bartram, ``Adaptive control: stability, convergence, and robustness,'' 1990.

\bibitem{mrac2}
P.~Ioannou and B.~Fidan, {\em Adaptive control tutorial}.
\newblock SIAM, 2006.

\bibitem{classMRAC}
K.~S. Narendra and A.~M. Annaswamy, {\em Stable adaptive systems}.
\newblock Courier Corporation, 2012.

\bibitem{BLF}
K.~P. Tee, S.~S. Ge, and E.~H. Tay, ``Barrier lyapunov functions for the control of output-constrained nonlinear systems,'' {\em Automatica}, vol.~45, no.~4, pp.~918--927, 2009.

\bibitem{BLF2}
Y.-J. Liu and S.~Tong, ``Barrier lyapunov functions-based adaptive control for a class of nonlinear pure-feedback systems with full state constraints,'' {\em Automatica}, vol.~64, pp.~70--75, 2016.

\bibitem{Lafflitto}
A.~L’Afflitto, ``Barrier lyapunov functions and constrained model reference adaptive control,'' {\em IEEE Control Systems Letters}, vol.~2, no.~3, pp.~441--446, 2018.

\bibitem{mpc11}
D.~Q. Mayne, J.~B. Rawlings, C.~V. Rao, and P.~O. Scokaert, ``Constrained model predictive control: Stability and optimality,'' {\em Automatica}, vol.~36, no.~6, pp.~789--814, 2000.

\bibitem{mpc12}
M.~V. Kothare, V.~Balakrishnan, and M.~Morari, ``Robust constrained model predictive control using linear matrix inequalities,'' {\em Automatica}, vol.~32, no.~10, pp.~1361--1379, 1996.

\bibitem{cbf}
A.~D. Ames, J.~W. Grizzle, and P.~Tabuada, ``Control barrier function based quadratic programs with application to adaptive cruise control,'' in {\em 53rd IEEE Conference on Decision and Control}, pp.~6271--6278, IEEE, 2014.

\bibitem{taylor2020adaptive}
A.~J. Taylor and A.~D. Ames, ``Adaptive safety with control barrier functions,'' in {\em 2020 American Control Conference (ACC)}, pp.~1399--1405, IEEE, 2020.

\bibitem{tvblf1}
K.~P. Tee, B.~Ren, and S.~S. Ge, ``Control of nonlinear systems with time-varying output constraints,'' {\em Automatica}, vol.~47, no.~11, pp.~2511--2516, 2011.

\bibitem{ppc1}
C.~P. Bechlioulis and G.~A. Rovithakis, ``Robust adaptive control of feedback linearizable mimo nonlinear systems with prescribed performance,'' {\em IEEE Transactions on Automatic Control}, vol.~53, no.~9, pp.~2090--2099, 2008.

\bibitem{ppc2}
C.~P. Bechlioulis and G.~A. Rovithakis, ``Prescribed performance adaptive control for multi-input multi-output affine in the control nonlinear systems,'' {\em IEEE Transactions on automatic control}, vol.~55, no.~5, pp.~1220--1226, 2010.

\bibitem{fc1}
T.~Berger, A.~Ilchmann, and E.~P. Ryan, ``Funnel control of nonlinear systems,'' {\em Mathematics of Control, Signals, and Systems}, vol.~33, pp.~151--194, 2021.

\bibitem{fc2}
A.~Hastir, J.~J. Winkin, and D.~Dochain, ``Funnel control for a class of nonlinear infinite-dimensional systems,'' {\em Automatica}, vol.~152, p.~110964, 2023.

\bibitem{ppc3}
F.~Fotiadis and G.~A. Rovithakis, ``Input-constrained prescribed performance control for high-order mimo uncertain nonlinear systems via reference modification,'' {\em IEEE Transactions on Automatic Control}, vol.~69, no.~5, pp.~3301--3308, 2023.

\bibitem{funnelsaturation1}
N.~Hopfe, A.~Ilchmann, and E.~P. Ryan, ``Funnel control with saturation: Linear mimo systems,'' {\em IEEE Transactions on Automatic Control}, vol.~55, no.~2, pp.~532--538, 2010.

\bibitem{fc3}
T.~Berger, ``Input-constrained funnel control of nonlinear systems,'' {\em IEEE Transactions on Automatic Control}, vol.~69, no.~8, pp.~5368--5382, 2024.

\bibitem{ghosh2022state}
P.~Ghosh and S.~Bhasin, ``State and input constrained model reference adaptive control,'' in {\em 2022 IEEE 61st Conference on Decision and Control (CDC)}, pp.~68--73, IEEE, 2022.

\bibitem{nussbaum}
Y.-J. Liu and S.~Tong, ``Barrier lyapunov functions for nussbaum gain adaptive control of full state constrained nonlinear systems,'' {\em Automatica}, vol.~76, pp.~143--152, 2017.

\bibitem{lavretsky2012robust}
E.~Lavretsky and K.~A. Wise, ``Robust adaptive control,'' in {\em Robust and adaptive control: With aerospace applications}, pp.~317--353, Springer, 2012.

\bibitem{lavretsky2011projection}
E.~Lavretsky and T.~E. Gibson, ``Projection operator in adaptive systems,'' {\em arXiv preprint arXiv:1112.4232}, 2011.

\bibitem{Sontag}
E.~D. Sontag, {\em Mathematical control theory: deterministic finite dimensional systems}, vol.~6.
\newblock Springer Science \& Business Media, 2013.

\bibitem{worthmann2020funnel}
K.~Worthmann, J.~Deutscher, and S.~Streif, ``A funnel control approach to output tracking for linear systems with input constraints,'' {\em Automatica}, vol.~118, p.~109022, 2020.

\bibitem{rajamani2012vehicle}
R.~Rajamani, {\em Vehicle Dynamics and Control}.
\newblock New York: Springer, 2~ed., 2012.

\end{thebibliography}

\end{document}